# Innovations in High- and Ultra-precision Machining

*E. Uhlmann, M. Polte, T. Hocke and J. Tschöpel*
Institute of Machine Tools and Factory Management, Technische Universität Berlin, Germany

**Abstract**
Modern precision manufacturing faces the challenge of integrating accuracy requirements into a framework of agile and sustainable production technologies. This development leads to numerous further challenges, affecting almost all areas of precision manufacturing industry. To overcome those challenges, this article presents promising technologies and innovative solution concepts from the field of temperature measurement in the cutting zone, process design for advanced materials, the development of reconfigurable machine tools, the optimization of the temperature behaviour of high-speed spindles and the use of edge computing in machine tools. Using the presented solutions, modern precision machining can be made more future-oriented in terms of sustainability and resilience.

## 1    Introduction to High- and Ultra-precision Machining

This article highlights outstanding technological innovations that were developed in recent years in the field of milling and turning manufacturing processes. These processes, which are assigned to group 3.2 'Cutting with geometrically defined cutting edges' with the subgroups 3.2.1 (turning) and 3.2.3 (milling) within the DIN 8580:2022 [1], represent the most industry-relevant and prevalent manufacturing technologies. For this reason, a more detailed explanation of the remaining manufacturing processes in group 3.2 such as planing, broaching or sawing, is not provided. As part of this CERN ACCELERATOR SCHOOL, which addresses the design and manufacture of highly complex components for particle accelerators and their associated detectors, this article focuses on the relevant high-precision (HP) and ultra-precision (UP) processes. These technologies, which are collectively referred to as "micro-machining", differ significantly from macro-machining in terms of the much lower chip thickness $h_{cu}$ involved in the cutting process. While the chip thickness $h_{cu}$ is in the range of a few millimetres in conventional macro-machining processes, the chip thicknesses $h_{cu}$ in micro-machining are characterised by only a few micrometres. As a result, the value of the chip thickness $h_{cu}$ is approximately equal to or significantly smaller than the rounded cutting edge radius $r_\beta$. This limits the specific contact between the cutting edge of the applied cutting tools and the workpiece to the rounded cutting edge radius $r_\beta$ and results in cutting conditions that differ significantly from those of macro-machining [2]. Further informations are provided in Section 3. Process Technologies for Ductile-Regime Cutting of Brittle Materials'.

A primary distinction between high-precision and ultra-precision processes can be made using the achievable shape accuracies $a_s$ and the surface roughness values. Secondary distinctions can be determined by considering the used tools and cutting materials. However, it is not always possible to clearly distinguish between the two technologies and the specific conditions need to be considered on a case-by-case basis [3, 4]. Nevertheless, Table 1 provides an overview of possible distinguishing criteria.

Table 1: Possible differentiating factors of high- and ultra-precision machining [3, 4]

| Criteria | High-precision machining | Ultra-precision machining |
|---|---|---|
| Prevailing machining process | Turning, Milling, Grinding | Single-Point-Diamond Turning, Fly-Cutting, Turning, Lapping |
| Conventional used cutting material | Cemented carbide | Single crystal diamond or polycrystalline diamond |
| Achievable shape accuracies $a_s$ | $a_s > 1$ µm | $a_s < 1$ µm |
| Achievable surface roughness value Ra | Ra > 10 nm | Ra < 10 nm |

The use of the mentioned manufacturing processes to produce components for medical technology, aerospace, electronics and semiconductor industry requires in-depth knowledge of the respective processes. Different requirements are imposed on the machining process depending on the application such as the production of optical surface qualities or complex micro-structures as well as the applied materials. If the specified process boundary conditions for the application are not achieved by the selected machine and process technology, the quality of the component will be significantly impaired. Therefore, the following sections describe specific challenges in high- and ultra-precision machining and the corresponding dedicated solutions.

## 2 Temperature Measurement System for UP-Turning Processes

### 2.1 Background and Motivation

Single crystal diamond tools are primarily used for the machining of component surfaces with optical properties using UP-turning processes. This cutting material is characterised by its extremely high hardness H in combination with the possibility of producing exceptionally small rounded cutting edge radii $r_\beta$. The rounded cutting edge radius $r_\beta$ can be achieved in the range of 10 nm $\leq r_\beta \leq$ 50 nm using special lapping and polishing processes [5, 6]. Single-point diamond turning (SPDT), in which a defined point on the tool cutting edge is used for machining, can be used to produce surfaces showing a surface roughness value of Ra < 5 nm. The primary benefit of this approach is that the inaccuracies of the cutting edge, characterised by its waviness $w_t$, are not transferred to the finished surface as only a single point of the cutting edge is in contact with the workpiece. Over the past decades, this process was one of the most relevant manufacturing technologies for optical components with high requirements concerning the shape accuracies $a_s$ as well as surface roughness values. Based on significant advances in manufacturing methods for diamond turning tools, it was possible to achieve a strong reduction in waviness $w_t$ along the cutting edge. According to the current state-of-the-art, it is feasible to obtain a waviness $w_t$ in the range of 50 nm < $w_t$ < 2 µm along the cutting edge [7, 8]. Consequently, the entire cutting edge of the diamond tool can be utilized for the machining process. For this reason, SPDT is now only used in application specific cases.

However, during the machining process, temperature-induced wear occurs on the single crystal diamonds [9, 10]. This results in a limited tool life $t_t$, which restricts the economic viability and efficiency of diamond turning processes. In order to increase the economic efficiency and the understanding of the wear effects in ultra-precision turning, the characterisation and interpretation of the cutting edge temperatures $\vartheta_{Ce}$, generated within the contact area between the turning tool and the workpiece, are of utmost importance. According to the state-of-the-art, there are no precise temperature methods available for online temperature monitoring during the process at the cutting edge with regard to the requirements for resolution accuracy $a_R$, response time $t_R$ and accessibility to the cutting edge. Temperature measurement methods such as thermocouples [11-13], pyrometers and thermographic cameras [14-16] were developed but due to their inadequate measurement characteristics, they are not suitable for highly sensitive temperature measurements in ultra-precision processes. For this purpose, a dedicated cutting edge temperature measurement system based on ion-implanted boron-doped single crystal diamonds as a highly sensitive temperature sensor for ultra-precision turning as part of several research projects was developed.

## 2.2 Methodology

The process is based on the use of single-crystal diamonds that are doped with boron-atoms and therefore provide an electrically conductive path along the cutting edge [17]. The integration of impurity atoms into the lattice structure of a diamond can convert it from an electrical insulator into an electrically conductive material. The chemical element boron was selected as the doping element as its atoms are characterized by three outer electrons, which, compared to a carbon atom from the diamond lattice with 4 valence electrons, results in a missing electron in the valence band of the material. This vacancy can be described as a free electron hole, which leads to electrical conductivity $\kappa$. This type of doping is referred to as p-doping [18]. For the realization of the doping process, ion-implantation was chosen to create the specific boron-doped path. The ion-implantation is a widely used method for doping, whereby chemical impurities in a solid-state material are caused by the targeted implantation of ions. This method offers the advantage of enabling very precise control of the doping density $\rho_{bor}$ and the spatial resolution $x_{bor}$ of the implanted area. This made it possible to reduce the distance between the boron-doped path and the cutting edge to $d_t = 90$ μm. Accordingly, the temperature $\vartheta$ is measured very close to the cutting edge, which significantly increased the measurement accuracy $a_R$ as well as response times $t_R$. To identify the doping characteristics in terms of a suitable doping density $\rho_{bor}$, doping width $d_{wid}$, doping length $d_{len}$ and doping depth $d_{dep}$, extensive investigations were conducted. According to the obtained influences of the doping characteristics on the electrical resistance characteristics, a doping area with a doping width of $d_{wid} = 44$ μm, a doping length of $d_{wid} = 420$ μm and a doping density of $\rho_{bor} = 3E^{15}$ ions/cm$^2$ was selected. The basic structure of the system and the course of the boron-doped path are shown in Fig. 1. To validate this measuring system, the temperature-resistance characteristic curve of the boron-doped diamonds was determined. For this purpose, a high-precision wafer-therm chuck system type SP 74A from ERS ELEKTRONIK GMBH, Hagen, was used, which allows a defined setting of the temperature $\vartheta$ with an accuracy of $a_\vartheta = 0.1$ K. By measuring the electrical resistance R while simultaneously varying the temperature $\vartheta$ of the wafer-therm chuck system at constant increments of $\vartheta_I = 1$ °C, the correlation between the electrical resistance $R_{el}$ and the temperature in ion-implanted boron-doped single crystal diamonds could be demonstrated for the first time. The measured correlation is shown in Fig. 2.

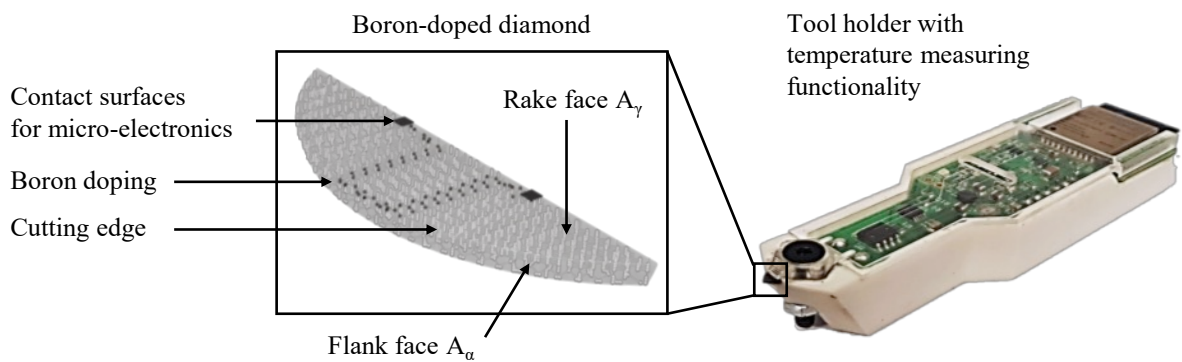

**Fig. 1**: Ion-implanted boron-doped single crystal diamond as a temperature sensor for UP-turning [17]

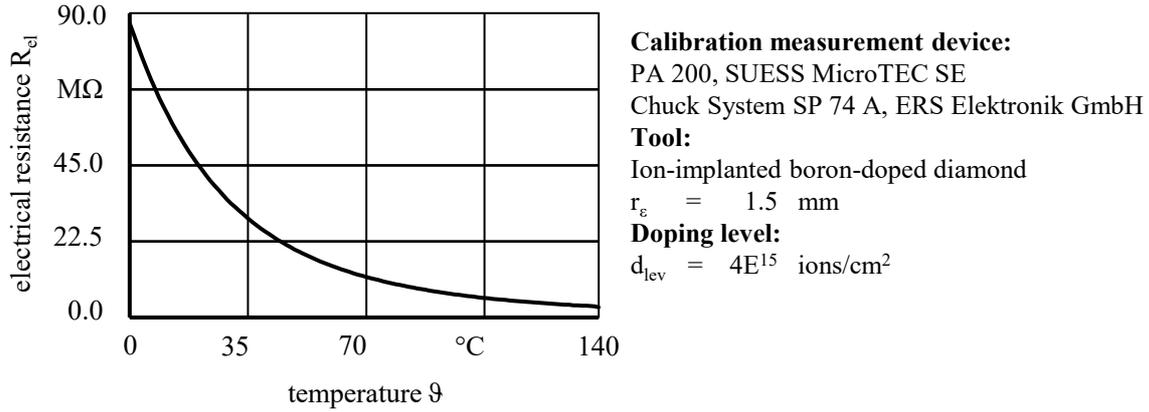

**Fig. 2**: Calibration of the ion-implanted boron-doped single crystal diamond [17]

## 2.3 Results

Previous research work showed that the developed temperature measurement system provides a measurement accuracy of $a_R = 0.4$ K and response times in the range of 420 ms $\leq t_R \leq$ 440 ms during the turning of polymethylmethacrylate (PMMA). This was determined for both finishing and roughing processes using a rake angle in the range of -30° $\leq \gamma_0 \leq$ 0° [17]. This study provides empirical evidences of the stability and reliability of the developed temperature measurement system for use in the field of ultra-precision machining compared to the current state-of-the-art.

The developed temperature measurement system was then employed to investigate the chip formation process for the material PMMA. It was demonstrated that the cutting edge temperature $\vartheta_{Ce}$ exerts a substantial influence on the chip formation. The results indicated that the feed f and chip thickness $h_{cu}$ significantly influenced temperature-related chip formation and the associated change in material structure. High-resolution scanning electron microscope images of the resulting chips were used to identify the change in the surface structure of the PMMA material. The chips exhibited a distinct melting of the chip surface when the glass transition temperature $\vartheta_G$ of the PMMA material was surpassed. By comparing the surface structure of the resulting chips with the measured cutting edge temperature $\vartheta_{Ce}$, conclusions could be drawn about the distribution of the heat energy Q deposited on the tool or the chip/air/workpiece. This analysis revealed a clear distinction between finishing parameters and roughing parameters [17].

In addition, the temperature development in the cutting zone was investigated for different materials. An overview of the measured cutting temperatures $\vartheta_C$ is shown in Fig. 3. In order to ensure direct comparability of the cutting temperature $\vartheta_C$ in all investigations, the same process parameters were utilized. The findings indicated that the cutting temperature $\vartheta_{Ce}$ for metallic materials are considerably higher than those for plastics. This phenomenon can be attributed to the presence of chemical bonds within the materials. In the case of plastics, van der Waals bonds, hydrogen bonds and dipole-dipole bonds predominate. In contrast, the structure of metals is predicated on metallic bonds. These exhibit considerably greater binding forces in comparison with the chemical bonds of polymeric materials. This indicates that a significantly higher energy is required for the chip formation process. As can be seen in Fig. 3, a cutting temperature of 36.87 °C $\leq \vartheta_C \leq$ 42.28 °C was observed with metallic materials, whereas the cutting temperature with plastic materials ranged within 22.01 °C $\leq \vartheta_C \leq$ 25.34 °C. The varying cutting temperature $\vartheta_C$ observed across these material groups can be attributed to the energy demands associated with material separation. This energy is found to be largely correlated with the tensile strength $\sigma_T$ of the material. Consequently, materials exhibiting the highest tensile strengths $\sigma_T$ also possess the highest cutting temperature $\vartheta_C$.

Based on the recorded cutting temperatures $\vartheta_C$ for different materials, the influences of the aforementioned process parameters on the temperature development within the cutting zone could be identified. These dependencies are fully described in POLTE ET. AL. [17]. These findings enable a well-founded prediction of the expected cutting edge temperatures $\vartheta_{Ce}$ as a function of the process parameters and, consequently, also statements about the expected tool wear. This allows the expected tool life $t_t$ to be determined much more precisely, which can significantly increase the usability and efficiency of diamond cutting processes in industrial applications.

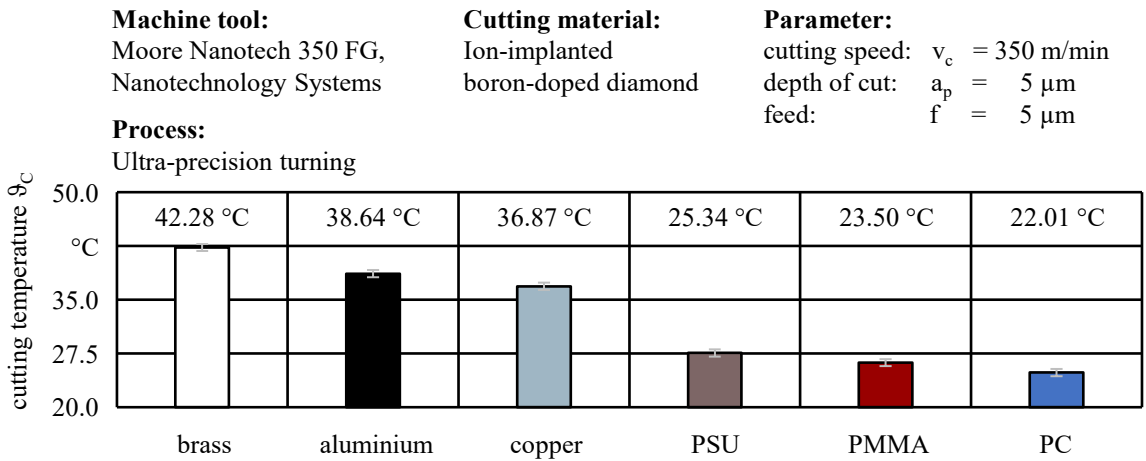

**Fig. 3**: Comparison of temperature levels of different materials [17]

In summary, the developed cutting-edge temperature measurement system based on ion-implanted boron-doped single crystal diamonds as a temperature sensor provides fundamental insights into the detailed characterization of the temperature development in correlation with chip formation mechanisms during UP-turning. Building on this, further research will extend the understanding of material-specific temperature behaviour with a particular focus on its relationship to the wear behaviour of single-crystal diamonds.

## 3 Process Technologies for Ductile-Regime Cutting of Brittle Materials

### 3.1 Background and Motivation

The production of surfaces of optical quality is one of the primary goals of the UP-machining. While this goal was achieved primarily for ductile materials such as copper or aluminium through the development of optimized cutting tools and the identification of suitable process parameters, the production of optical surfaces on components made of brittle-hard materials such as single-crystal germanium or silicon continues to represent a key challenge in the state-of-the-art [19]. The reason for this lies in the chip formation during the machining of brittle and brittle-hard materials. Due to the low plastic deformability of the materials, the chip formation is characterized by the development and enlargement of micro-cracks. These cracks occur in the surface layer of the material due to the application of a defined hydrostatic pressure stress state with the tool [20]. According to POLTE [POL16], the achievement of this complex stress state is possible during machining, when utilizing an effective negative rake angle $\gamma_{eff}$ (see Fig. 4). Using an effective negative rake angle $\gamma_{eff}$ significantly reduces the formation of cracks in the machined surface. The effective negative rake angle $\gamma_{eff}$ can be realised by taking the ratio of the chip thickness to the rounded cutting edge radius of $h_{cu} / r_\beta \leq 1$ into account. A plastic deformation of the workpiece material takes place when the shear stress $T_{max}$ reaches a critical material-dependent value within the shear plane. However, material breakage arises when the effective tensile stress $\sigma_t$ attains a material-dependent critical value.

It was shown that the formation of cracks can be significantly influenced by the ratio of the chip thickness to the rounded cutting edge radius of $h_{cu} / r_\beta \leq 1$, which are defined by the process parameters and the applied cutting tool. If the maximum chip thickness $h_{cu,max}$ is too large, cracks form near the surface that extend deeper than the machined surface with the used depth of cut $a_p$, thereby reducing the surface quality of the machined area. This correlation is shown in Fig. 5.

This fundamental cutting mechanism was already scientifically described in early research works [2, 21-23].

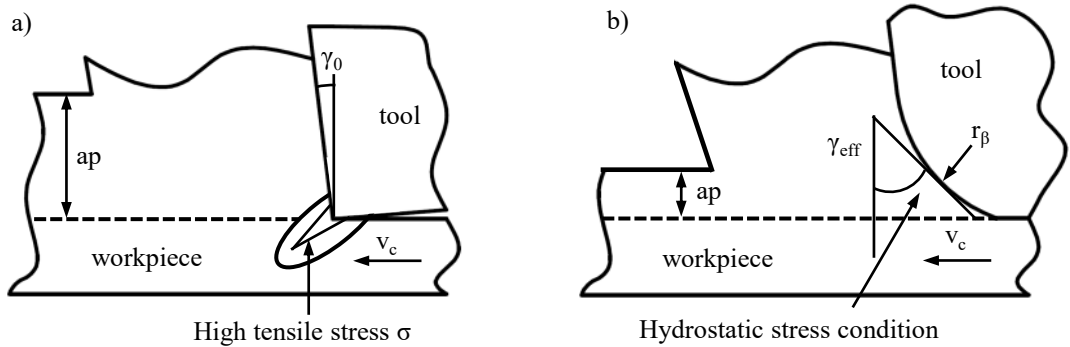

**Fig. 4**: a) Negative rake angle $\gamma_0$ through the macroscopic tool geometry and b) effective negative rake angle $\gamma_{eff}$ through the microscopic tool geometry, adapted from YAN [24].

By selecting appropriate process parameters and taking all relevant process conditions into account, such as the machine tool vibration effects, the application of coolant or the tool wear condition, it is possible to achieve a hydrostatic process state in which the brittle material exhibits ductile machining properties. This process state is generally referred to as 'ductile-regime cutting' and should always be aimed for in order to achieve increased surface qualities [19].

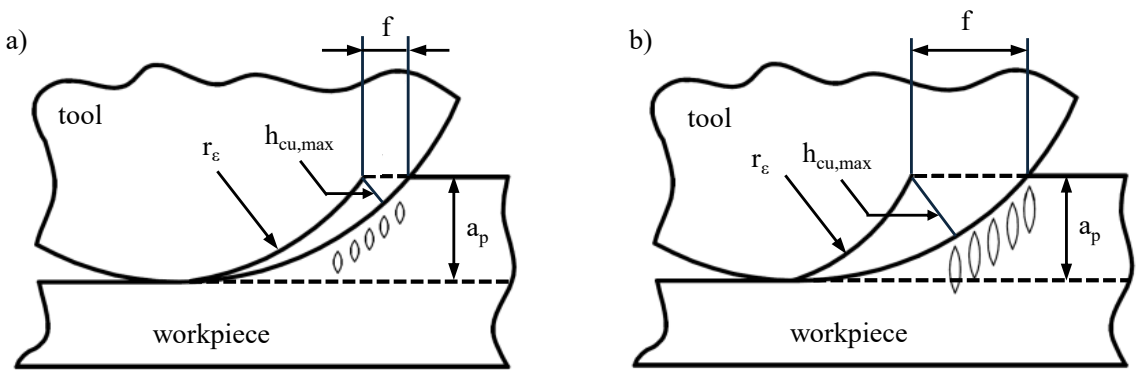

**Fig. 5**: Crack propagation in dependence on the feed f; a) microcracks above the machined surface; b) microcracks below the machined surface, adapted from BLACKLEY [22].

## 3.2 Methodology

Due to the strong dependence of the process parameters for achieving a 'ductile-regime cutting' on the properties of the material, these need to be re-identified for each workpiece material. In recent times, binderless cemented carbide was observed as a suitable workpiece material to gain considerable industrial relevance. This material offers promising properties for the use as a mould insert in injection moulding processes or cutting or punching tools. This is due to the significantly lower mass fraction of the cobalt binder phase with approximately $C_{CO} = 0.5$ % compared to conventional cemented carbide with $C_{CO} = 11 - 13$ %, which results in a significantly improved wear behaviour. Extensive machining tests were carried out to provide a suitable process technology for a ductile machining of this workpiece material.

In order to gain knowledge about the specific minimum chip thickness $h_{cu,min}$, scratch tests were carried out. In these experimental investigations, were the tool was translated in a linear direction over the workpiece, the cutting depth $a_p$ as well as the feed f was continuously increased. The resulting grooves were evaluated using scanning electron microscope images, which enabled the ductile areas to be identified [25]. Innovative cutting tools made of nanopolycrystalline diamond were applied to machine this workpiece material, which compared to conventional single crystal diamonds exhibit a significantly improved cutting edge stability as well as improved wear behaviour.

### 3.3 Results

The results of these investigations confirmed that ductile machining characteristics can be achieved by selecting suitable process parameters. A dominant ductile-regime cutting was observed for a ratio of the chip thickness to the rounded cutting edge radius $0.269 \leq (h_{cu} / r_\beta) \leq 0.781$. This process condition was established by setting the feed within $3\ \mu m \leq f \leq 8\ \mu m$ and the depth of cut within $3\ \mu m \leq a_p \leq 8\ \mu m$. The findings further demonstrated, that the feed f exerts a substantial influence on the development and enlargement of micro-cracks within the machined surface, reducing its quality. The identification of the process parameters enabled the manufacturing of surfaces with a surface roughness value Ra = 11.0 nm and a maximum profile height Rz = 71.7 nm (see Fig. 6). A feed f = 3 µm, a cutting depth of $a_p$ = 3 µm and a cutting speed $v_c$ = 50 m/min were selected for this machining process.

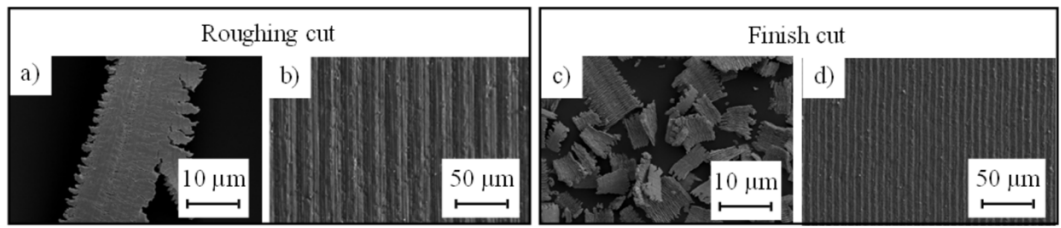

**Fig. 6**: Machining results from a turning process in binderless cemented carbide with a) and b) chips and surface texture within the roughing process as well as c) and d) chips and surface texture within finishing process [25, 26].

The results showed that suitable process parameters enable the machining of demanding hard and brittle materials such as binderless cemented carbide. This technological development is economically significant, for example, in the manufacturing of punching tools made from this material, as it can significantly increase their tool life $t_t$ and consequently their shape accuracy $a_S$ over a high number of punching cycles $n_p$. This leads to considerable cost savings and enables a more economical production in the area of precision manufacturing.

## 4 Thermoelectrically Controlled Milling Spindle

### 4.1 Background and Motivation

In many cases, the use of milling tools with a tool diameter of D < 1 mm is necessary for producing complex geometric micro-structures. Due to the small tool diameter D, high rotational speeds n are required in order to achieve sufficient cutting speeds $v_c$. High-speed spindles are used for this purpose, which can achieve rotational speeds of n > 200,000 1/min. However, high rotational speeds n cause significant friction effects within the spindle bearings, resulting in thermally induced axial elongation of the spindle shaft $\Delta l_s$. The axial elongation $\Delta l_s$ exerts a direct influence on the position of the tool centre point (TCP) and thus on the achievable machining accuracy $a_m$. Numerous solutions already exist to reduce and compensate the axial elongation $\Delta l_s$. For example, heat pipes were integrated into the spindle bearing area, allowing a more precise temperature control of the bearings [27, 28].

The disadvantage of this approach is its increased complexity, which is a consequence of the supplementary implementation of the heat pipes within the spindle structure. Due to the resulting increase in manufacturing costs, such spindle systems are rarely used in industrial applications.

A further approach to reduce the axial elongation of the spindle $\Delta l_s$ involves the variation of the spindle power $P_{sp}$ in such a way that a minimum of heat energy Q is introduced into the spindle bearings [29]. It is important to note that, with this approach, a complete compensation of the axial elongation of the spindle $\Delta l_s$ can only be guaranteed within a limited operating range of the spindle, regarding the spindle power $P_{sp}$. This can be regarded as a significant disadvantage of the solution. In addition, the use of air-bearing high-frequency spindles was also proposed as a possible solution. In the absence of solid contact between the rotor and stator, friction effects and axial elongation of the spindle shaft $\Delta l_s$ are not observed [30]. The primary disadvantage associated with this solution is the necessity to utilize compressed air with a pressure of $p_a \leq 30$ bar. The provision of compressed air results in a significant increase in operating costs with a negative environmental impact compared to a conventional ball-bearing spindle. As a result, the aforementioned approaches are not yet established in commercially available machine tools on a large scale. The current state-of-the-art is to operate the tool spindle at idle speed $n_{id}$ until a thermally stable state is reached [31-33]. During these warm-up phases, no machining process can take place, which significantly reduces the overall efficiency $E_{sum}$ of the machine tool.

### 4.2 Methodology

To solve this technological challenge, UHLMANN developed a thermoelectrically tempered spindle [34, 35]. This approach is based on the integration of Peltier elements between the ball bearings and the cooling channels of the spindle (see Fig. 7). Peltier elements utilize the inverse Seebeck effect. This enables them to precisely control the heat flow rate $\dot{Q}$ depending on the electric current $I_{Pelt}$ flowing through them. As a result, the temperature $\vartheta_{sp}$ of the spindle bearings can be controlled in a targeted manner. In cooperation with NAKANISHI JAEGER GMBH, Ober-Mörlen, Germany, a functional prototype based on the Z62 spindle was developed.

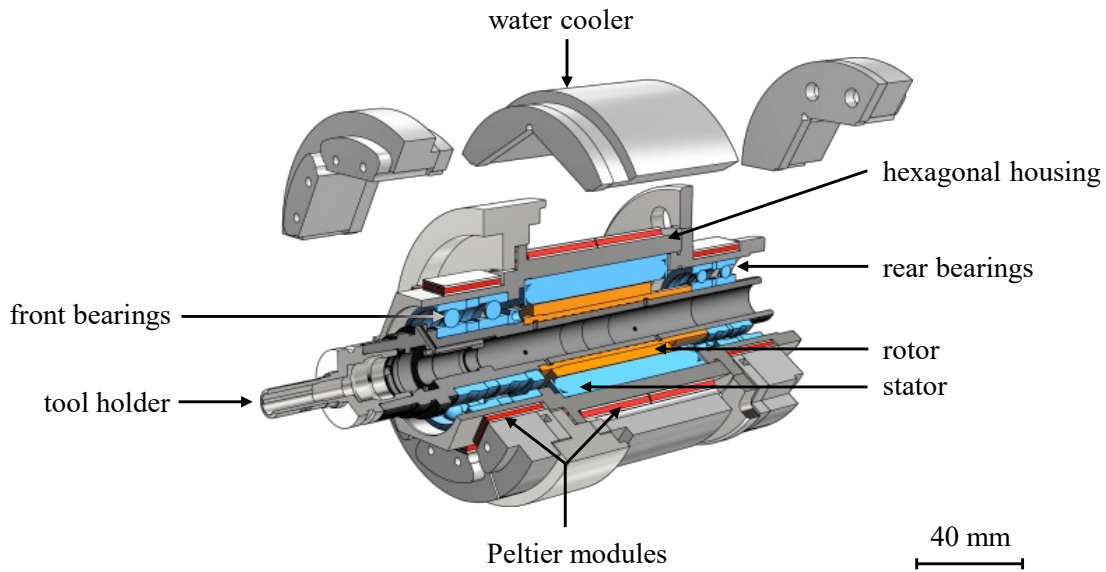

**Fig. 7**: Integration of Peltier elements into the thermoelectrically tempered spindle Z62 from the manufacturer NAKANISHI JAEGER GMBH, Ober-Mörlen, Germany.

## 4.3 Results

The functionality described above offers two possible applications. Firstly, the heat flow rate $\dot{Q}$ can be significantly reduced during the warm-up phase, allowing the spindle shaft and bearings to heat up more rapidly and thus reach a thermally stable state much faster. Experimental investigations showed that the warm-up phase, after the spindle stopped for a time of t = 300 s, can be shortened from $t_w$ = 343 s with the reference spindle to $t_w$ = 71 s with the thermoelectric spindle. This corresponds to a reduction of 86 % (see Fig. 8) [36]. Another possible application is the targeted control of the heat flow rate $\dot{Q}$ between the heat source (spindle bearing) and the heat sink (cooling channels) during the machining process, which allows varying thermal loads and, consequently, varying axial elongations of the spindle shaft $\Delta l_s$ to be compensated. This can significantly increase the resulting machining accuracy $a_m$.

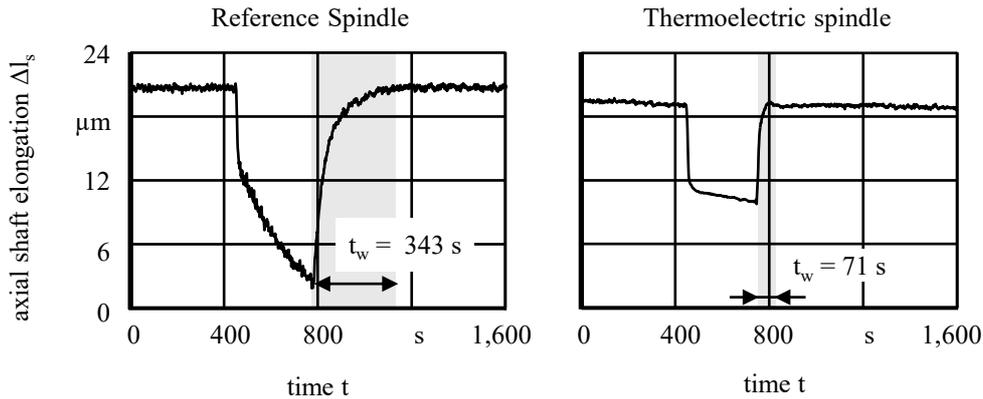

**Fig. 8**: Comparison of warm-up durations $t_w$ of different spindle types according to UHLMANN [36].

## 5 Modularized and Reconfigurable Machine Tool

### 5.1 Background and Motivation

The entire industrial production sector is undergoing continuous change. One of the most prominent developments in this context concerns the increasing individualization of products and the shortening of product life cycles. Consequently, manufacturing processes need to be adapted to the varying products at shorter intervals in order to accommodate the shorter product development phases and increasing variety of variants. In order to ensure economical production with standard delivery times and manufacturing costs, it is essential that the relevant production systems provide a high degree of flexibility. Machine tools are of paramount importance in this regard, since their design necessitates their construction and optimisation for defined applications. For example, this applies to the weight and dimensions of the workpieces. To solve this technological challenge, KOREN [37] introduced the concept of the "reconfigurable machine system" in the 1990s. Subsequent to this, a considerable number of research projects were initiated on the basis of this concept [38, 39]. For instance, WULFSBERG ET AL. [40] developed a miniaturised milling system that features interchangeable drive elements as well as an adaptable basic structure. A reconfigurable machine tool with a significantly larger machining area was further presented by LORENZER [41]. However, this does not permit the adjustment of the machine frame. Despite the existence of numerous academic prototypes, only a few reconfigurable machine tools were manufactured in large quantities [42]. Consequently, there is currently no machine tool available that allows comprehensive configuration of the machine frame, all drive elements as well as the spindle system. Nevertheless, the establishment of such a system is imperative to ensure high productivity even with small lot sizes and widely varying component shapes.

## 5.2 Methodology

PEUKERT [43] and UHLMANN [44] implemented the concept of a 'reconfigurable machine system' using the example of a reconfigurable milling machine. Fig. 9a shows the modular machine system in an example configuration as a portal milling machine. Hexagonal topology-optimized building blocks, which can be connected to each other by means of screw connections, served as the basis for the machine frame (see Fig. 9b). These passive modules were made from cast steel type GS 52 (material number: 1.0552 [45]). To increase the achievable machining accuracy $a_m$, the passive components were supplemented with active components.

For this purpose, an active module was designed that can compensate for static displacements $\Delta x_{stat}$ of the machine frame. These are caused by high workpiece masses $m_{ws}$ or local heating $\Delta \vartheta_l$. This compensation module exhibits the same dimensions as the passive building blocks and can therefore be integrated at any point within the machine structure. The functionality of this module is based on the thermal elongation $\Delta l$ of an integrated aluminium body, which can be measured and controlled using strain gauges. The structure of this module is shown in Fig. 9c. The linear expansion of the module can be adjusted with high precision in the range of $-48\ \mu m \leq x_a \leq 24\ \mu m$ [43].

Additionally, an active workpiece clamping module was developed to compensate for dynamic vibration effects, which can compensate for dynamic displacements of the TCP by deliberately displacing the workpiece [46]. This module consists of a complex solid-state joint that restricts the degrees of freedom of the workpiece clamping to translational movements in the x- and y-directions as well as two correspondingly aligned piezo-active actuators. The structure of this system is shown in Fig. 9d. In addition, an active damping module was developed and qualified, which also exhibits the same dimensions and can therefore be integrated into the machine setup as required.

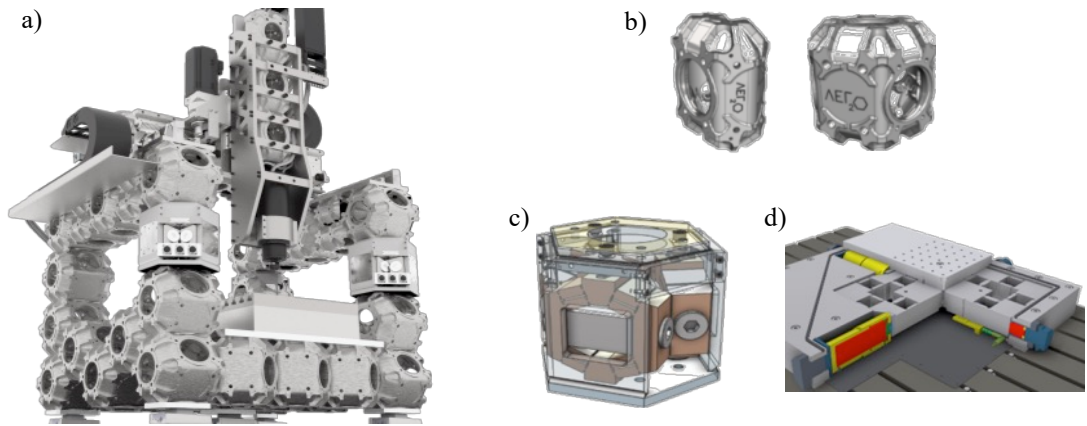

**Fig. 9**: Reconfigurable machine tool with a) exemplary configuration as a portal-type milling machine; b) passive building blocks; c) active module for compensating static displacement $\Delta x_{stat}$ and d) active workpiece clamping system for enhancing the dynamic positioning accuracy $a_{pos.}$

## 5.3 Results

Extensive experimental investigations were carried out to validate the prototypical machine system. As part of these investigations, stability lobe diagrams were determined, among other things, which show the limits of a stable machining process [44]. For this purpose, the influence of the spindle speed n and the cutting depth $a_p$ on the stability of a milling process was determined. A high-strength aluminium alloy of type EN AW 7075 was used as the workpiece material. Based on the measured accelerations $a_m$ at the milling spindle and the workpiece clamping, the limit range of a stable machining process could be determined with increasing spindle speed n and increasing cutting depth $a_p$. With the help of these investigations, the maximum achievable material removal rate $Q_W$ could be determined. The stable milling process was limited by a maximum cutting depth of $a_{p,lim} = 1.4$ mm. Furthermore, the positioning accuracy $a_a$ of the axes was measured.

For example, it was found that the positioning accuracy of the Y-axis in the range from 0 mm ≤ y ≤ 300 mm amounts to $a_y ≤ 3$ μm. The results of the qualification of the modular milling machine showed that the concept of a 'reconfigurable machine system' can be used to build a milling machine that offers performance characteristics comparable to those of commercially available machines. Consequently, this machine tool makes a significant contribution to the flexibility of production systems. This approach ensures cost-effective production, even in the context of a wide variety of products and short delivery times.

# 6 Development of Machine-Leaning Models Using Edge Computing

## 6.1 Background and Motivation

Monitoring the process stability in machining processes is an important prerequisite for complying with component tolerance specifications. Process instabilities in the form of chatter vibrations significantly reduce the resulting surface quality and shorten the tool life $t_t$. Stability lobe diagrams, which must be determined in extensive experimental investigations, are used to examine these vibration effects [47]. To simplify these time-consuming experiments, they can also be carried out automatically. However, this requires access to the machine tool control system in order to be able to adjust the motion control according to the chatter vibrations that occur during the experiments.

## 6.2 Methodology

The required interface to the machine control system was implemented by UHLMANN [48] using an industrial edge device of the type Simantik IPC427E from SIEMENS AG, Erlangen, Germany. With the help of this edge device, high-frequency machine data can be read out at sampling interval of $\Delta t_{pos}$ = 2 ms, as well as low-frequency data at a sampling interval of $\Delta tp_{os}$ = 100 ms. The high-frequency data originates directly from the machine's position control loop and can therefore be used to adjust the machine's motion control for the investigation of chatter vibrations. For this purpose, the power $P_a$, the current $I_a$, the drive torque $M_a$ and accelerations $a_a$ of the drive axes were recorded using the build-in sensors of the machine tool. Furthermore, by comparing the target positions $x_{tar}$ with the recorded actual positions $x_{act}$, information about the control difference $\Delta x$ of the corresponding machine axes could be obtained. The utilization of these signals facilitates the allocation of the machining process to either a stable or unstable process state. It is evident that, due to the substantial number of input data being employed, the occurrence of the unstable process state can be detected in an entirely automated manner. The schematic structure of the described data acquisition is shown in Fig. 10.

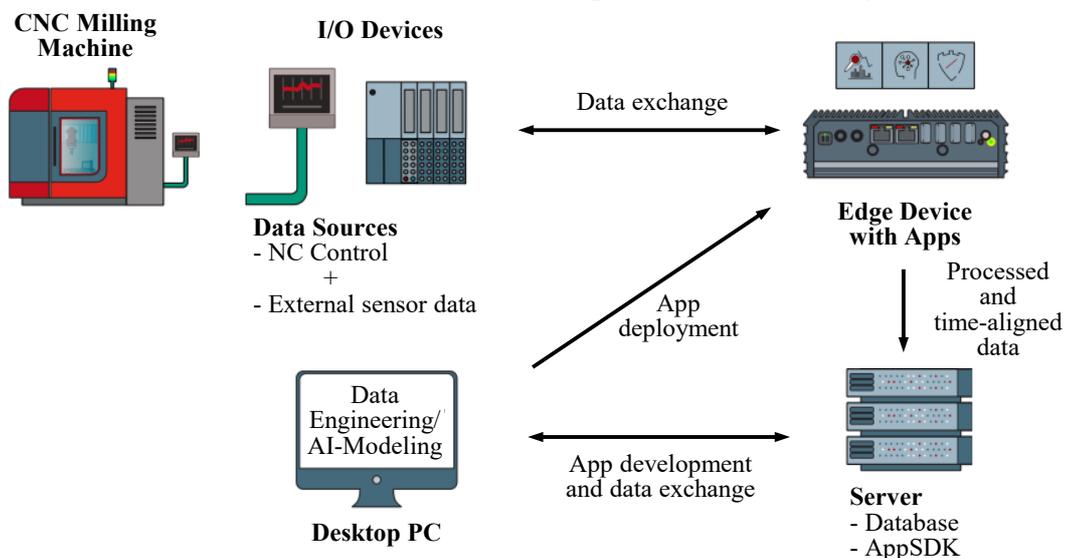

**Fig. 10**: Data acquisition from machine tools using edge computing [48]

## 6.3 Results

This approach enables the stability limits of a machining process to be determined automatically. This resulted in a substantial reduction in the time and personnel required. With the help of this technology, the occurrence of chatter vibrations in the machining process can be recorded automatically and evaluated in detail.

The gained process knowledge in this way enables a significant optimization of the machining parameters with regard to an increased process efficiency in the form of a reduced energy consumption or a decreased machining time without reducing the stability of the process. Furthermore, the ability to automatically extract and provide large amounts of labelled data sets from the machine tool is particularly important for artificial intelligence applications, as the availability of data for training AI models is a key challenge. In this context, there are also numerous other fields of application for industrial edge devices, such as predictive maintenance or adaptive control for various machining processes. Moreover, this approach has the potential to make a substantial contribution to the optimization of the dynamic properties of machine tools, obviating the necessity for extensive and consequently cost-intensive external sensor technology.